\documentclass[aps,prfluids,showpacs,notitlepage,superscriptaddress,longbibliography]{revtex4-2}
\usepackage{graphicx}
\usepackage{hyperref}
\usepackage[utf8]{inputenc}

\usepackage{amsmath,latexsym,amssymb,amsfonts}
\usepackage{bm}
\newcommand{\vect}[1]{\boldsymbol{#1}}

\begin{document}

\title{Optimal swimmer can be puller, pusher, or neutral depending on the shape}

\author{Abdallah Daddi-Moussa-Ider}
\affiliation{Institut f\"{u}r Theoretische Physik II: Weiche Materie, Heinrich-Heine-Universit\"{a}t D\"{u}sseldorf, 40225 D\"{u}sseldorf, Germany}
\author{Babak Nasouri}
\affiliation{Max Planck Institute for Dynamics and Self-Organization (MPIDS), 37077 G\"ottingen, Germany}
\author{Andrej Vilfan}
\email[E-mail: ]{andrej.vilfan@ds.mpg.de}
\affiliation{Max Planck Institute for Dynamics and Self-Organization (MPIDS), 37077 G\"ottingen, Germany}
\affiliation{Jo\v{z}ef Stefan Institute, 1000 Ljubljana, Slovenia}
\author{Ramin Golestanian}
\affiliation{Max Planck Institute for Dynamics and Self-Organization (MPIDS), 37077 G\"ottingen, Germany}
\affiliation{Rudolf Peierls Centre for Theoretical Physics, University of Oxford, Oxford OX1 3PU, United Kingdom}

\begin{abstract}
The ability of microswimmers to deploy optimal propulsion strategies is of paramount importance for their locomotory performance and survival at low Reynolds numbers. Although for perfectly spherical swimmers minimum dissipation requires a neutral type swimming, any departure from the spherical shape may lead the swimmer to adopt a new propulsion strategy, namely those of puller- or pusher-type swimming. In this study, by using the minimum dissipation theorem for microswimmers, we determine the flow field of an optimal nearly spherical swimmer, and show that indeed depending on the shape profile, the optimal swimmer can be a puller, pusher, or neutral. Using an asymptotic approach, we find that amongst all the modes of the shape function, only the third mode determines to leading order the swimming type of the optimal swimmer.
\end{abstract}

\maketitle

\section{Introduction}
\label{sec:intro}

An active particle (or microswimmer), be it a living cell or a synthetic swimmer, converts the internal or ambient free energy into work as it moves through a viscous fluid \citep{Lauga2009,bechinger16,gompper2020}.
From a broad hydrodynamic perspective, the physics behind the propulsion of an active swimmer can be divided into two parts: the \emph{inner} problem which concerns the generation of the propulsive thrust, and the \emph{outer} problem which focuses on how swimmers interact with their neighbouring environment through altering their surrounding fluid. While in the former accounting for the details of the mechanism behind the impetus of each specific swimmer is essential (for example through cilia \citep{blake1974} or a phoretic mechanism \citep{golestanian05,nasouri2020}), in the latter, one can use a generic approach to describe the flow field induced by the swimmer \citep{kim1991,lauga2016stresslets,nasouri2018higher}. 
Specifically for self-propelling axisymmetric swimmers, this generic approach classifies the swimmers into three groups of pushers, pullers, and neutrals, often referred to as the microswimming types \citep{Underhill.Graham2008,Lauga2009}. This categorization, which stems from the far-field description of the motion of a particle in a viscous fluid, relies on the fact that self-propulsion is force- and torque-free, and so the leading-order flow field induced by an active swimmer can be solely described by a symmetric force dipole, i.e. stresslet \citep{batchelor1970stress}. Based on the strength of this force dipole, a swimmer is a puller when it generates the impetus from its front end, a pusher when the thrust originates from the rear end, and is neutral when this strength is zero. One example of pusher-type microswimmers include \emph{E.\ coli} bacteria that utilize bundles of rotating helical filaments in their rear \citep{berke08}, or sperm cells that propel themselves by propagating a wave along a flexible flagellum. An example of puller-type microswimmers is \emph{Chlamydomonas reinhardtii} that pulls in the fluid in front of it with a pair of flagella beating in a breaststroke-like fashion \citep{kantsler2013ciliary}. \emph{Volvox}, a multicellular colony of green algae, is a neutral swimmer \citep{drescher2009,Pedley.Goldstein2016}, whereas \emph{Paramecium} is a weak pusher \citep{Zhang.Jung2015}. 

Swimmers of different type behave differently in interacting with their surroundings. 
For instance, unlike puller-like swimmers, pushers can be hydrodynamically trapped by nearby obstacles, or other pusher swimmers \citep{berke08,spagnolie2015, daddi18jpcm, sprenger2020towards}. The stresslet further determines the intensity of fluid stirring in suspensions of swimmers \citep{Lin.Childress2011}. 
Although the effect of swimming type on the interaction of each swimmer with other swimmers/boundaries has been well explored, their energetic implications are yet to be fully understood. For surface-driven spherical swimmers, it has been shown that the viscous dissipation of neutral swimmers is minimal compared to that of pushers and pullers, and so neutral swimmers are often considered as the optimal type \citep{michelin2010}. However,
the innate question of whether this statement holds when the swimmer does not possess a perfect spherical shape, remains largely unanswered.
This is the question we address in this study.

{
The question of energetic efficiency and optimal propulsion, i.e. minimizing the dissipation while maintaining the swimming speed or equivalently maximizing the swimming speed while maintaining the dissipation, is a long-standing problem.
Earlier theoretical works focused on the optimal locomotion of flagellated micro-organisms \citep{pironneau1974optimal, lighthill1975}.
In particular, the optimal shape of a periodically actuated planar flagellum deforming via a travelling wave has been derived computationally \citep{lauga2013shape}, and shown to agree well with the waveform assumed by sperm cells of marine organisms.
The optimal swimming strokes and self-propulsion efficiencies of spherical and cylindrical bodies undergoing small deformation with respect to a reference shape has also been investigated \citep{shapere1987self, shapere1989efficiencies}.
Further studies considered the full optimization problem for simple mechanically-actuated model microswimmers \citep{alouges2007,Nasouri.Golestanian2019}.
}

{
Generally, the quest for the optimal propulsion strategy requires both the solution of the inner and the outer problem.
}
Swimming efficiency of ciliated microswimmers can be directly determined numerically \citep{Ito.Ishikawa2019,Omori.Ishikawa2020}, but it is more common to use a coarse grained approach, namely to separately calculate the dissipation in the propulsive layer and then replace this layer with an effective slip velocity when determining the external flow \citep{Keller.Wu1977,osterman2011,vilfan2012,sabass2010}. A fundamental limit on swimming efficiency can be obtained by finding the slip profile that minimizes the external dissipation for a given swimming speed.
For spherical swimmers, by using the classical squirmer model of \citet{lighthill1952} and \citet{blake1971spherical}, one can show that the contribution of the second mode of squirming (which characterizes the stresslet) to the dissipated power can only be positive. Because the swimming speed for spherical squirmers is independent of this second mode, we can conclude that minimizing the dissipation requires the second mode to be zero, thereby making the optimal swimmer a neutral one \citep{Blake1973}. However, such a simple decomposition of contributions cannot be achieved for non-spherical swimmers, and so the correlation between the dipole coefficient and the dissipation is not clearly known. Recently, using the boundary element method and numerical optimization, \citet{guo2021optimal} showed on some example shapes that when the swimmer body is not front-aft symmetric, pushers or pullers can be more efficient than neutral swimmers. In this study, we systematically investigate the relation between the stresslet and the shape of nearly-spherical optimal swimmers. By employing the recently derived minimum dissipation theorem \citep{Nasouri.Golestanian2021}, we circumvent the nonlinear optimization problem and arrive at the flow field for the optimal swimmer using the flow fields of two auxiliary passive problems. We remarkably find that the stresslet of an optimal swimmer is solely function of the third Legendre mode describing the shape of the swimmer, and so depending on the value (or sign) of this mode, the optimal swimmer can be a pusher, puller, or neutral.

\section{The problem statement}

\begin{figure}
  \centering
  \includegraphics[width=0.6\textwidth]{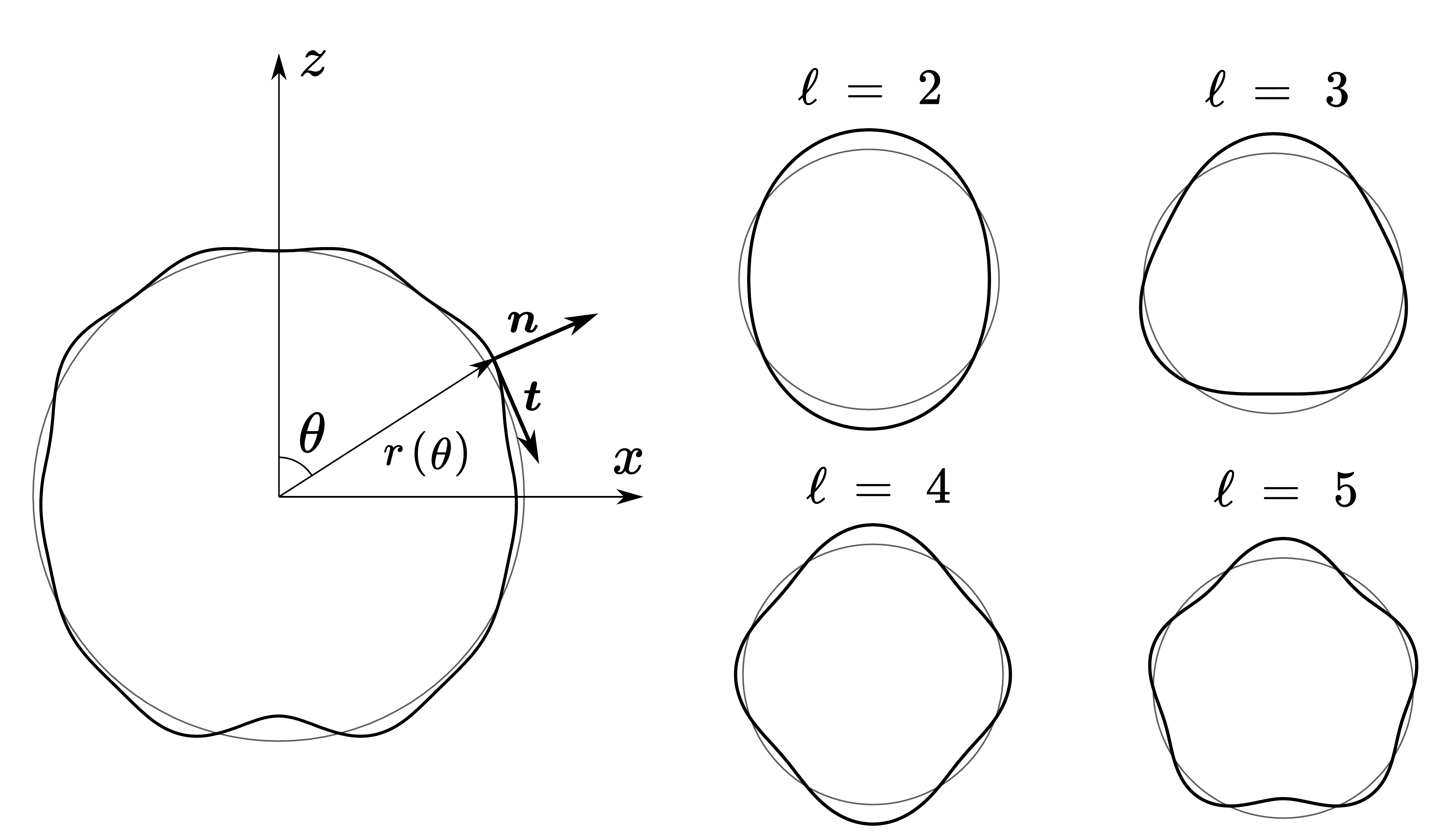}
	\caption{{Schematic of the nearly-spherical swimmer described in equation~\eqref{shape_function}. The panels on the right show the isolated contribution of the first four modes in the shape function. Black lines show the perturbed shape and the gray lines illustrate the reference unperturbed sphere.  
}}
	\label{fig:schematic}
\end{figure}

In this study, our aim is to determine whether an optimal nearly spherical swimmer is a puller, pusher, or neutral. To this end, we consider a swimming body of axisymmetric shape moving with a steady velocity~$V_\mathrm{A}\bm{e}_z$, where $\bm{e}_z$ is a unit vector representing the axis of symmetry.
We parameterize the surface of the swimming object in axisymmetric spherical coordinates by
\begin{align}
\label{shape_function}
r(\theta)= a \left[ 1 + \sum_{\ell=1}^\infty \alpha_\ell P_\ell(\cos\theta) \right],
\end{align}
where $a$ denotes the radius of the undeformed sphere, $\theta$ represents the polar angle with respect to $\bm{e}_z$ and
$P_\ell$ is the Legendre polynomial of degree~$\ell$ (see figure~\ref{fig:schematic}). We assume $\alpha_\ell\ll 1$, thus the particle possesses a nearly spherical shape. Note that since the first mode merely implies body translation and does not indicate any departure from the spherical shape, we set $\alpha_1=0$.

At the small scales of microswimmers, viscous forces dominate inertial forces, and the flow is governed by the Stokes equations $\boldsymbol{\nabla} \cdot \boldsymbol{\sigma} = \boldsymbol{0}$ and $\boldsymbol{\nabla} \cdot \vect{v} = 0$ with $\vect{v}$ denoting the flow field, $\boldsymbol{\sigma}=-p \vect{I} +\mu\left(  \boldsymbol{\nabla} \vect{v}  + \boldsymbol{\nabla} \vect{v}^\top\right)$ the stress field, and $p$ the pressure field. The swimmer is surface-driven and its active mechanism induces an effective tangential slip-velocity $\vect{v}^s$ on its surface, which imposes the boundary condition on the fluid velocity in the co-moving frame $\vect{v}=\vect{v}^s$. The slip profile determines the swimming velocity $V_A$ through a relationship that can be derived from the Lorentz reciprocal theorem \citep{stone1996}. The dissipated power is given by $P=-\int \vect{v}^s \cdot \vect{\sigma} \cdot \vect{n}\, dS$. 
We consider the swimmer to be optimal, thus this slip profile minimizes the viscous dissipation $P$, while maintaining the swimming speed $V_\mathrm{A}$.

{
We should note that the present analytical description of microswimmers applies exclusively to non-deformable active swimmers of nearly-spherical shape.
Prime examples of these swimmers include a broad class of ciliated microorganisms or synthetic microswimmers that achieve locomotion via a thin slip layer (e.g. self-phoretic mechanisms).
}

As discussed earlier, the far-field flow generated by the force- and torque-free motion of a microswimmer has the form (to the leading order) $\vect v(\vect x)= -(3 / (8\pi \mu)) (\vect x \cdot \vect S \cdot \vect x) \, \vect x/r^5$ and is characterized by the stresslet $\vect S$. Here, since the motion is axisymmetric, the stresslet takes the simple form of
\begin{equation}
  \vect{S} = 8\pi \mu a^2 V_A\, \beta
  \left( \vect{e}_z  \vect{e}_z - \tfrac{1}{3} \, \vect{I} \right),
\label{eq:stresslet}  
\end{equation}
where $\beta$ is the dimensionless dipole coefficient \citep{batchelor1970stress,nasouri2018higher}. Under this definition, the sign of $\beta$ determines the swimming type such that $\beta < 0$ holds for pushers, $\beta > 0$ for pullers, and $\beta=0$ indicates neutral swimming. Thus, to determine the swimming type of an optimal nearly spherical swimmer, we need to find the relation between $\beta$ and $\alpha_\ell$.

Conventionally, finding the flow field surrounding an optimal swimmer requires extensive optimization schemes, which are often implemented by the means of computational tools. Here, we alternatively apply a fundamental theorem that sets the lower bound on the energy dissipation of a self-propelled active microswimmer of arbitrary shape \citep{Nasouri.Golestanian2021}. It states that the motion of an active swimmer with minimal dissipation can be conveniently expressed as a linear superposition of two passive bodies of the same shape satisfying no-slip and perfect-slip boundary conditions at their surfaces, respectively. This theorem relies on the fact that perfect-slip bodies require the least dissipation for motion, suggesting that a swimmer with a \textit{similar} slip profile will be more efficient. A superposition with the no-slip problem is needed to obtain a force-free flow around an active swimmer (see \citet{Nasouri.Golestanian2021} for the details of the derivation). Specifically, defining $\bm{v}_A$ as the flow field induced by the motion of the optimal swimmer, this theorem dictates
\begin{align}
\label{theorem}
\bm{v}_A=\bm{v}_\text{PS}-\bm{v}_\text{NS}
\end{align}
where $\bm{v}_\text{PS}$ is the flow field due to the motion of a passive perfect-slip body of the same shape translating with speed $V_\text{PS}=[R_\text{NS}/(R_\text{NS}-R_\text{PS})]{V}_A$, and $\bm{v}_\text{NS}$ is the flow field of its no-slip counterpart moving with speed $V_\text{NS}=[R_\text{PS}/(R_\text{NS}-R_\text{PS})] {V}_A$, with ${R}_\text{NS}$ and ${R}_\text{PS}$ being the translational drag coefficients for the no-slip and the perfect-slip body, respectively.

Accordingly, by the means of this theorem, the optimization problem is reduced to finding the flow fields of two passive systems (henceforth referred to using `PS' and `NS') and their corresponding drag coefficients. Following an asymptotic approach, we will prove that the dipole coefficient takes a particularly simple expression and can solely be expressed in terms of the third Legendre mode as
\begin{equation}
	\beta = \frac{27}{14} \, \alpha_3 \, .  \label{centralResult}
\end{equation}
Based on this, the nearly spherical optimal swimmer is classified as a pusher when $\alpha_3 < 0$, puller when $\alpha_3 > 0$ and neutral if $\alpha_3 = 0$.

\section{Solution of the passive problem}

As discussed earlier, to find the flow field of the optimal active swimmer, we only need to determine the flow fields around a passive body of the same shape, once with a no-slip and once with a perfect-slip boundary condition. 

Recalling that the particle is nearly spherical (i.e., $\alpha_\ell \ll 1$), we use an asymptotic approach in finding the flow fields, and expand all entities in terms of surface modes. At the zeroth order (denoted by `(0)'), we recover the flow fields due to the passive motion of a spherical particle with no-slip and perfect-slip boundary conditions. The first-order correction (denoted by `(1)') will then be due to the surface departure from the spherical shape, and so based on the linearity of the field equations, must be a linear superposition of the surface modes, e.g. $\vect{v}=\vect{v}^{(0)}+\sum_\ell \alpha_\ell \vect{v}^{(1)}_\ell$. In what follows, we find the zeroth- and first-order flow fields for both the NS and PS problems, by applying the Lamb's solution at each order separately. 

We should also account for the correction to the surface normal vector at the first order.  At the zeroth order we have $\vect{n}^{(0)} = \vect{e}_r$, and the departure from spherical shape leads to 
$\vect{n}^{(1)} = -\sum_\ell \alpha_\ell P^1_\ell (\cos\theta) \, \vect{e}_\theta$ where $P_\ell^1(\cos\theta)=dP_\ell(\cos\theta)/d\theta$ is the associate Legendre polynomial of the first order. The tangent vector is given by $\vect{t}^{(0)} = \vect{e}_\theta$ and $\vect{t}^{(1)} = \sum_\ell \alpha_\ell P^1_\ell (\cos\theta) \, \vect{e}_r$.

\subsection{No-slip problem}

The solution for the flow past a nearly spherical body with a no-slip boundary is discussed in \citet{happel1983}. In the following, we derive the flow field in a form that will be convenient for the solution of the active problem in the next section. Due to linearity of the problem, we only need to solve the flow field for a single mode of surface deformation (e.g. $\alpha_\ell$), and the complete solution will be achieved by linear superposition of all modes.

In the co-moving frame of reference, the no-slip boundary condition requires vanishing velocities at the deformed surface of the object such that
\begin{equation}
	\vect{v}_\text{NS} = \vect{0} \qquad \text{at} \qquad r(\theta) =a\left[ 1 + \alpha_\ell P_\ell(\cos\theta)\right] \, .
\end{equation}
This condition can be expanded perturbatively to linear order in $\alpha_\ell$  as
\begin{equation}
  \left. \vect{v}^{(0)} + \alpha_\ell \left( \vect{v}^{(1)} +
      a \, \frac{\partial \vect{v}^{(0)}}{\partial r} \,  P_\ell(\cos\theta) \right)
  \right|_{r=a} = \bm{0} \, . 
\label{eq:bcns2}  
\end{equation}
To find the Stokes flow that satisfies the above boundary condition, along with the condition $\vect{v}=-V_{\text{NS}} \, \vect{e}_z$ at $r\to \infty$, we use Lamb's general solution in spherical coordinates as an ansatz \citep{happel1983}. For axisymmetric problems, it simplifies to 
	\begin{subequations} \label{velocityComponents}
		\begin{align}
			\frac{v_r}{V_\text{NS}} &= -\cos\theta +
			\sum_{n=1}^{\infty}
			\frac{n+1}{2} \left( {n} \,  A_n - 2{B_n} \left(\frac{a}{r}\right)^{2} \right) \left(\frac{a}{r}\right)^n P_n (\cos\theta)  \, , \\
			\frac{v_\theta}{V_\text{NS}} &=
			\sin\theta +
			\sum_{n=1}^{\infty}
			\left( - \frac{n-2}{2} \,  A_n + B_n \left(\frac{a}{r}\right)^{2}  \right) \left(\frac{a}{r}\right)^n  {P^1_n (\cos\theta)}  \, ,
		\end{align}
	\end{subequations}
where $A_n$ and $B_n$ are series coefficients that must be determined from the boundary conditions.

The solution for the zeroth-order problem corresponding to an undeformed sphere can readily be obtained by imposing $v_r^{(0)} = 0$ and $v_\theta^{(0)} = 0$ at $r=a$.
This leads us to
$A_1^{(0)} = {3/2}$, $B_1^{(0)}= {1/4}$, and $A_n^{(0)} = B_n^{(0)}= 0$ for $n \ge 2$. The zeroth order flow field
\begin{equation}
	\frac{v_r^{(0)}}{V_\mathrm{NS}} = -\frac{1}{2} \left( 2-\frac{3a}{r} + \frac{a^3}{r^3} \right) \cos\theta \, , \qquad
	\frac{v_\theta^{(0)}}{V_\mathrm{NS}} = \frac{1}{4} \left( 4-\frac{3a}{r} - \frac{a^3}{r^3} \right) \sin\theta \, ,
\end{equation}
represents the well known flow past a no-slip sphere \citep{happel1983}. 

The boundary condition for the first-order problem \eqref{eq:bcns2} then reads $\vect{v}^{(1)} = -a \, (\partial \vect{v}^{(0)}/\partial r) \, P_\ell(\cos\theta)$ at $r=a$.
By noting that $\partial v_r^{(0)}/\partial r = 0$ at $r=a$, we find upon using appropriate orthogonality relations that only the series coefficients of order $n=\ell \pm 1$ have non-zero values. 
Specifically, we find
$A_{\ell-1}^{(1)}=-A_{\ell+1}^{(1)}=-(3/2)/(2\ell+1)$,  $B_{\ell-1}^{(1)}=-(3/4)(\ell-1)/(2\ell+1)$, and  $B_{\ell+1}^{(1)}=(3/4)(\ell+1)/(2\ell+1)$.
The first-order correction to the flow can be evaluated by inserting these coefficients into the generic solution given in \eqref{velocityComponents}. Examples of flow patterns for the first three deformation modes are shown in the left column of figure~\ref{fig:flowFields}.

The drag force exerted on an object is always determined by force monopole as $F_\text{D} = -4\pi\mu a A_1 V_\text{NS}$.
Accordingly, the translational drag coefficient for an approximate sphere only depends on the zeroth and second Legendre modes and can be written as \citep{happel1983}
\begin{equation}
  \frac{{R}_\mathrm{NS}}{6\pi\mu a}
  = 1 - \frac{1}{5} \, \alpha_2 \, .
  \label{eq:drag-coeff-NS}
\end{equation}

\subsection{Perfect-slip problem}

For the perfect-slip boundary condition, the impermeability and vanishing tangential stress need to be satisfied at the surface of the approximate sphere,
\begin{equation}
	\vect{v}_\text{PS} \cdot \vect{n} = 0,  \quad \text{and} \quad \bm{t} \cdot
	\boldsymbol{\sigma}_\text{PS} \cdot \vect{n}={0}, \quad \text{at} \quad r(\theta) = a\left[1 + \alpha_\ell P_\ell(\cos\theta)\right] \, .
\end{equation}
A Taylor expansion up to linear order in $\alpha_\ell$ leads to
\begin{subequations}
	\begin{align}
		\left.  v_r^{(0)}
		+ \alpha_\ell \left( v_r^{(1)} + v_\theta^{(0)} P_\ell^1(\cos\theta)
		+ a \, \frac{\partial v_r^{(0)}}{\partial r} \, P_\ell(\cos\theta)
		\right) \right|_{r = a} &= 0 \, ,  \\ 
		\left. \sigma_{r\theta}^{(0)} \
		+ \alpha_\ell \left(
		\sigma_{r\theta}^{(1)}
		+  \left( \sigma_{rr}^{(0)} - \sigma_{\theta\theta}^{(0)} \right)P_\ell^1(\cos\theta)
		+  a \, \frac{\partial \sigma_{r\theta}^{(0)}}{\partial r} \, P_\ell(\cos\theta)
		 \right)  \right|_{r = a}
		 &= 0 \, .  
	\end{align}
 \end{subequations}
Again, we solve the flow problem using Lamb's solution \eqref{velocityComponents} and determine the coefficients $A_n$ and $B_n$ that satisfy the above conditions. The solution for the zeroth-order problem corresponding to an undeformed sphere is obtained by requiring $v_r^{(0)} = 0$ and $\sigma_{r\theta}^{(0)} = 0$, which readily leads us to $A_1^{(0)} = 1$, $B_1^{(0)} = 0$, and $A_n^{(0)}= B_n^{(0)}= 0$ for $n \ge 2$. Thus, at the zeroth order we have
\begin{equation}
	\frac{v_r^{(0)}}{V_\mathrm{PS}} = -\left( 1 - \frac{a}{r} \right) \cos\theta \, , \qquad
	\frac{v_\theta^{(0)}}{V_\mathrm{PS}} = \frac{1}{2} \left( 2 - \frac{a}{r} \right)\sin\theta 
\end{equation}
which, as expected, is the flow past a spherical air bubble \citep{happel1983}.

\begin{figure}
  \centering
  \includegraphics[width=0.65\textwidth]{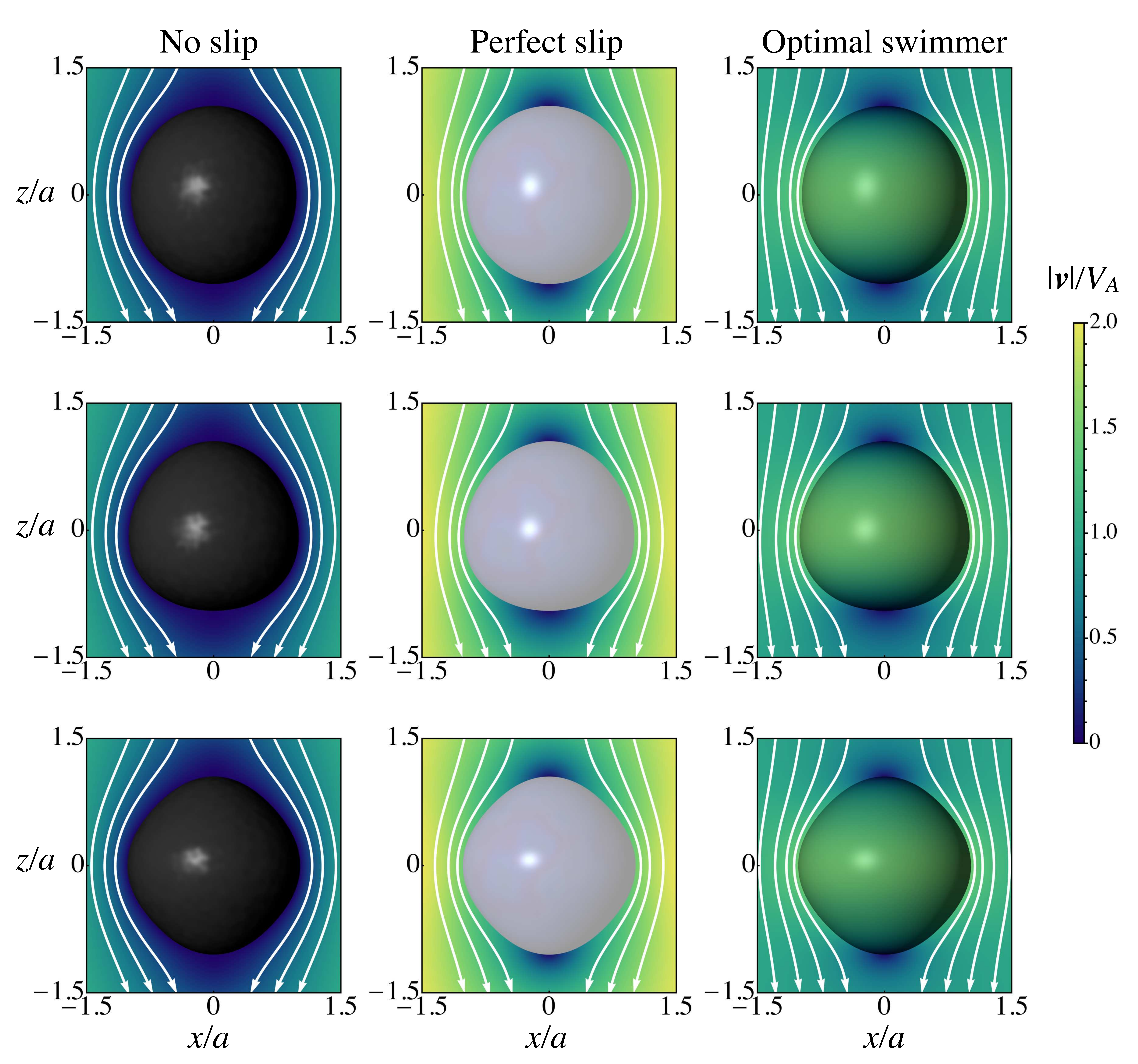}
	\caption{
	Streamlines around a slightly deformed sphere with no-slip (left column), perfect-slip (middle column), and optimal active swimmer (right column) in the co-moving frame. Each row shows one deformation mode with the amplitude $\alpha_\ell=0.05$ for $\ell=2$ (top row), $\ell=3$ (middle row), and $\ell=4$ (bottom row). The colour indicates the fluid velocity, scaled by the speed of the active swimmer.
}
	\label{fig:flowFields}
\end{figure}

Proceeding to the first order, noting that $\sigma_{r\theta}^{(0)} = \sigma_{\theta\theta}^{(0)} = 0$ everywhere in the fluid domain, we, again, find that all the terms except $n =\ell \pm 1$ are zero. The first-order coefficients due to the effect of $\alpha_\ell$ are thereby found
\begin{subequations}
	\begin{align}
		A_{\ell-1}^{(1)}&=-\cfrac{(\ell+1)(\ell+2)}{(2\ell-1)(2\ell+1)} \, ,
		& A_{\ell+1}^{(1)}&=\cfrac{\ell^2+\ell+3}{(2\ell+3)(2\ell+1)} \, ,\\
		B_{\ell-1}^{(1)}&=- \cfrac{(\ell-1)(\ell^2+\ell+3)}{2(2\ell-1)(2\ell+1)} \, ,
		& B_{\ell+1}^{(1)}&= \cfrac{\ell(\ell-1)(\ell+1)}{2(2\ell+3)(2\ell+1)} \, .
	\end{align}
\end{subequations}
These coefficients determine the first-order solution for the flow field with the perfect-slip boundary condition (figure~\ref{fig:flowFields}, middle column). From the drag force $F_\text{D} = -4\pi\mu a A_1 V_\text{PS}$, we determine the drag coefficient as
\begin{equation}
	\frac{{R}_\mathrm{PS}}{4\pi\mu a}
	= 1 - \frac{4}{5} \, \alpha_2 \, . \label{Drag-coeff-PS}
\end{equation}
The result is consistent with the calculation for an ellipsoidal particle, where only the deformation mode $\ell=2$ is present \citep{chang2009translation}, but has a broader validity, as it shows that deformation modes beyond the second do not influence the drag coefficient in linear order.

\section{Optimal active swimmer}
Having derived the solutions of the flow problems for no-slip and perfect-slip boundary conditions, we next make use of these solutions to construct the flow field induced by a self-propelling active microswimmer with minimum dissipation, i.e. the optimal swimmer. As shown in Eq.~\eqref{theorem}, the flow field surrounding the optimal swimmer can be reconstructed by a linear superposition of the flow fields of the no-slip and perfect-slip problems, weighted by a specific combination of their drag coefficients.

\subsection{Stresslet of the optimal microswimmer}

We first evaluate the stresslet of the optimal swimmer and its dipole coefficient. Since both passive flows are expanded in terms of Lamb's solution, their superposition, too, has the same form. A comparison between the flow field in \eqref{velocityComponents} and the definition of the stresslet \eqref{eq:stresslet} shows that only the coefficient $A_2$ contributes to the stresslet.
{
Specifically, the dipolar contribution to the flow field which decays as $r^{-2}$ reads $ \frac{\boldsymbol{v}}{V} = \frac{3}{2} A_2 \left( 3\cos^2\theta -1 \right) \left( \frac{a}{r} \right)^2 \boldsymbol{e}_r$, indicating that the dipole coefficient must be $\beta = -(3/2) A_2$.  Note that $A_2^{(1)}  \ne 0$ only for $\ell \in \{ 1,3 \}$ and here we have set $\alpha_1=0$, so
in the perturbative expansion the dipole coefficient evaluates to }
\begin{equation}
\label{result1}
	{\beta}
	= -\frac{3}{2} \, \mathcal{A}_2^{(1)} \alpha_3 \, ,
\end{equation}
with
\begin{align}
  \mathcal{A}_2^{(1)}=\frac{R_\text{NS}}{R_\text{NS}-R_\text{PS}} \left[ A_2^{(1)}\right]_\text{PS}-\frac{R_\text{PS}}{R_\text{NS}-R_\text{PS}}\left[ A_2^{(1)}\right]_\text{NS}
  \label{eq:a2superpositon}
\end{align}
being the corresponding coefficient of the active swimmer, expressed in terms of those of the NS and PS problems.

From Eq.~\eqref{result1}, one can see that the corrections to the drag coefficients $R_\text{NS}$ and $R_\text{PS}$ do not have any contributions to $\beta$ in the leading order. Equation \eqref{eq:a2superpositon} can therefore be evaluated with the drag coefficients of spherical particles. Remarkably, the dipole coefficient, to the leading order, only depends on the third Legendre mode of the shape function ($\alpha_3$), and other modes have no contribution.
Inserting the values of $A_2^{(1)}$ from the NS and PS calculations into Eq.~\eqref{result1}, we finally arrive at our final solution given in Eq.~\eqref{centralResult}.

\subsection{Flow field of the optimal microswimmer}

\begin{figure}
  \centering
  \includegraphics[width=0.8\textwidth]{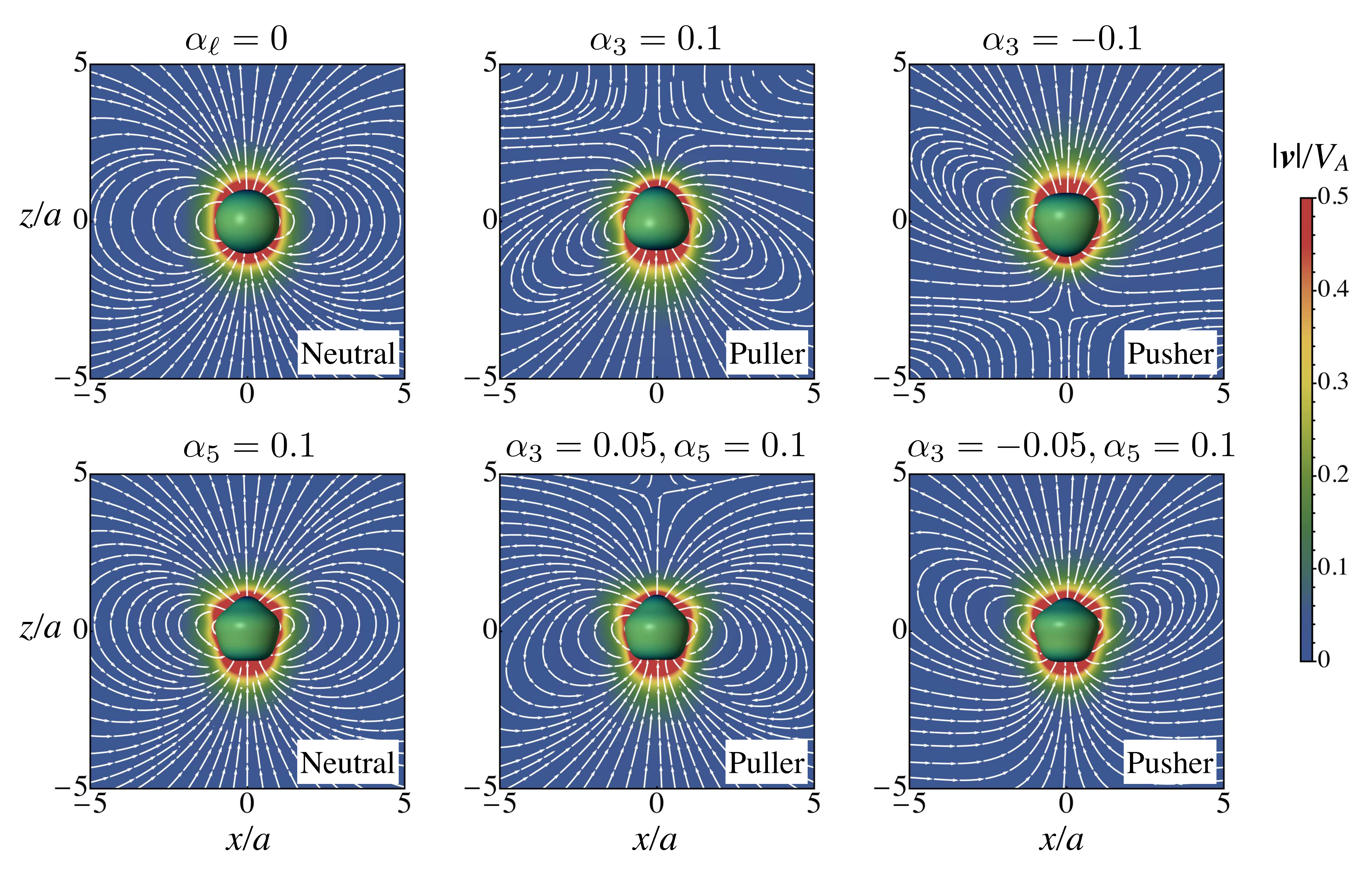}
	\caption{
	Streamlines in laboratory frame of optimal swimmers of various shapes. The nonzero surface modes for each swimmer are given at the top of its panel. The colours in the flow field indicate its velocity scaled by the swimming speed of the active particle. The swimmer surface colours represent the slip velocity as in figure~\ref{fig:flowFields}. 
}
	\label{fig:labframe}
\end{figure}

The full velocity field induced by the optimal active microswimmer (figure~\ref{fig:flowFields}, right column) can be obtained up to the linear order in deformation amplitudes by evaluating all coefficients in the same way as shown in Eq.~\eqref{eq:a2superpositon}. Thereby, the drag coefficients $R_\text{NS}$ and $R_\text{PS}$ need to be evaluated to linear order, as given in Eqs.~\eqref{eq:drag-coeff-NS} and \eqref{Drag-coeff-PS}. In figure~\ref{fig:labframe}, the flow fields for some nearly spherical optimal swimmers are shown in the laboratory frame.

\section{Conclusions}

In this study we analyzed the swimming type of nearly spherical optimal swimmers. We applied the minimum dissipation theorem \citep{Nasouri.Golestanian2021} to determine the  flow field of the optimal swimmer and to show that the dipole coefficient (or the strength of stresslet) only depends to leading order on the third mode of the shape function. Thus, depending on the sign of this mode, the optimal swimmer is a puller (when positive), pusher (when negative), or neutral (when zero). Using our results, one can determine the optimal swimming type for surface-driven nearly-spherical swimmers by simply describing the shape function in terms of the Legendre expansion and calculating the third mode. Our results can also be applied to phoretic particles which use their surface activity to gain propulsion. For instance, for a chemically-active particle, the slip velocity depends on the surface coating pattern which characterizes the chemical activity and mobility rates. For a given nearly-spherical phoretic particle, one can then use our results to determine whether that surface coating minimizes the viscous dissipation. In the hydrodynamically optimal case, the dipole coefficient follows from the shape as derived here. We should note that for optimizing phoretic particles, one should also account for the dissipation in the slip layer and the energetics of the chemical reaction \citep{sabass2010,sabass2012}, which is not considered here and can be a natural extension to this work.

Our derivation demonstrates how the recently proposed theorem can enable us to find a perturbative explicit solution to a problem that would otherwise hardly be analytically tractable. It is also possible to extend the presented results by accounting for the nonlinear effect of the quadratic and higher-order terms, in which case the contribution of other surface modes will be nonzero. Beyond that, one can use the methodology discussed here to evaluate the swimming type of any optimal swimmer of any arbitrary shape, provided the flow fields for the no-slip and perfect-slip problems are known.

\bibliography{bibliography}

\end{document}